\documentclass[11pt]{article}
\usepackage[margin=1.0in]{geometry}
\usepackage{amsmath}
\usepackage{mathrsfs}
\usepackage{amsfonts}
\usepackage{bbm}
\usepackage{fullpage}
\usepackage{graphicx}
\usepackage{moreverb}
\usepackage[ruled, vlined]{algorithm2e} 
\usepackage{cite}

\usepackage[font=small]{caption}

\newcommand{\argmax}{\operatornamewithlimits{argmax}}

\newcommand\numberthis{\addtocounter{equation}{1}\tag{\theequation}}

\author{Surojit Biswas$^{1}$, Meredith McDonald$^2$, Derek S. Lundberg$^2$, \\ Jeffery L. Dangl$^{2,3,4,5}$, Vladimir Jojic$^{6}$}
\title{Learning microbial interaction networks from metagenomic count data}
\date{}
\begin{document}



\maketitle
\noindent $^{1}$ Department of Statistics, University of North Carolina at Chapel Hill, Chapel Hill, NC 27599, USA \\
$^{2}$ Department of Biology, University of North Carolina at Chapel Hill, Chapel Hill, NC 27599, USA \\
$^{3}$ Howard Hughes Medical Institute, University of North Carolina, Chapel Hill, NC, 27599, USA \\
$^{4}$ Carolina Center for Genome Sciences, University of North Carolina, Chapel Hill, NC, 27599, USA \\
$^{5}$ Department of Immunology, University of North Carolina, Chapel Hill, NC, 27599, USA \\
$^{6}$ Department of Computer Science, University of North Carolina, Chapel Hill, NC, 27599, USA
\vspace{0mm}

 \begin{abstract}
Many microbes associate with higher eukaryotes and impact their vitality. In order to engineer microbiomes for host benefit, we must understand the rules of community assembly and maintenence, which in large part, demands an understanding of the direct interactions between community members. Toward this end, we've developed a Poisson-multivariate normal hierarchical model to learn direct interactions from the count-based output of standard metagenomics sequencing experiments. Our model controls for confounding predictors at the Poisson layer, and captures direct taxon-taxon interactions at the multivariate normal layer using an $\ell_1$ penalized precision matrix. We show in a synthetic experiment that our method handily outperforms state-of-the-art methods such as SparCC and the graphical lasso (glasso). In a real, \emph{in planta} perturbation experiment of a nine member bacterial community, we show our model, but not SparCC or glasso, correctly resolves a direct interaction structure among three community members that associate with \emph{Arabidopsis thaliana} roots. We conclude that our method provides a structured, accurate, and distributionally reasonable way of modeling correlated count based random variables and capturing direct interactions among them. \\
\\
\textbf{Availibility:} We have implemented our Poisson-multivariate normal hierarchical model in an R package named MInt. The package will be made available on CRAN.

 \end{abstract}

\section*{Introduction}
Microbes are the most diverse form of life on the planet. Many associate with higher eukaryotes, including humans and plants, and perform key metabolic functions that underpin host viability \cite{The2012, Lundberg2012}. Importantly, they coexist in these ecologies in various symbiotic relationships \cite{Konopka2009}.  Understanding the structure of their interaction networks may simplify the list of microbial targets that can be modulated for host benefit, or assembled into small artificial communities that are deliverable as probiotics.

Microbiomes can be measured by sequencing all host-associated 16S rRNA gene content. Because the 16S gene is a faithful phylogenetic marker, this approach readily reveals the taxonomic composition of the host metagenome \cite{Segata2013}. Given such sequencing experiments output an integral, non-negative number of sequencing reads, the final output for such an experiment can be summarized in a $n$-samples $\times$ $s$-taxa count table, $Y$, where $Y_{ij}$ denotes the number of reads that map taxon $j$ in sample $i$. It is assumed $Y_{ij}$ is proportional to taxon $j$'s true abundance in sample $i$. 

To study interrelationships between taxa, we require a method that transforms $Y$ into an undirected graph represented by a symmetric and weighted $o \times o$ adjacency matrix, $A$, where a non-zero entry in position $(i,j)$ indicates an association between taxon $i$ and taxon $j$. Correlation-based methods are a popular approach to achieve this end \cite{Faust2012, Friedman2012, Faust2012a}.  Nevertheless, correlated taxa need not directly interact if, for example, they are co-regulated by a third taxon. Gaussian graphical models remedy this concern by estimating a conditional independence network in which $A_{ij} = 0$ if and only if taxon $i$ and taxon $j$ are conditionally independent given all remaining taxa under consideration \cite{Meinshausen2006, Friedman2008, Wainwright2008}. However, they also assume the columns of $Y$ are normally distributed, which is unreasonable for a metagenomic sequencing experiment. Finally, neither correlation nor Gaussian graphical modeling offer a systematic way to control for confounding predictors, such as measured biological covariates (e.g. body site, or plant fraction), experimental replicate, sequencing plate, or sequencing depth. 

As baseline methods, we consider the commonly used correlation network method, SparCC \cite{Friedman2012}, and a state-of-the-art method for inferring Gaussian graphical models, the graphical lasso \cite{Friedman2008}. SparCC calculates an approximate linear correlation between taxa after taking the log-ratio of their abundances, and through an iterative procedure, prunes correlation edges that are not robust. In this way, it not only aims to produce a sparse network, but also avoids negative correlations between taxa that arise from data compositionality -- a common problem in metagenomics experiments, in which counts of taxa can only be interpreted relative to eachother, and not as absolute abundance measurements. Importantly, the authors point out that SparCC does not make any parametric assumptions.

The graphical lasso aims to construct a sparse graphical model, in which non-zero edges can be interpreted as direct interactions between taxa. Model inference proceeds by optimizing the likelihood of a standard multivariate normal distribution with respect to the precision matrix, subject to an $\ell_1$ constraint on each entry. The magnitude of this $\ell_1$ penalty controls the degree of sparsity, or equivalently, model parsimony.

In this work, we develop a Poisson-multivariate normal hierarchical model that can account for correlation structure among count-based random variables. Our model controls for confounding predictors at the Poisson layer, and captures direct taxon-taxon interactions at the multivariate normal layer using an $\ell_1$ penalized precision matrix.

\section*{Materials and Methods}
\subsection*{Preliminaries}
Let $n$, $p$, and $o$ denote the number of samples, number of predictors, and the number of response variables under consideration, respectively. Throughout this paper, response variables will be read counts of bacteria and will be referred to as such, though in practice, any count based random variable is relevant.  Let $Y$ be the $n \times o$ response matrix, where $Y_{ij}$ denotes the count of bacteria $j$ in sample $i$. Let $X$ be the $n \times p$ design matrix, where $X_{ij}$ denotes predictor $j$'s value for sample $i$. For a matrix $M$, we will use the notation $M_{:i}$ and $M_{i:}$ to index the entire $i^{th}$ column and row, respectively. The Frobenius norm of $M$ is defined to be $||M||_F = \sqrt{ \sum_i\sum_j M_{ij}^2}$.

\subsection*{The model}

We wish to model direct interaction relationships among bacteria measured in a metagenomic sequencing experiment while also controlling for the confounding biological and/or technical predictors encoded in $X$. Toward this end, we propose the following Poisson-multivariate normal hierarchical model. 
\begin{align*}
w &\sim \textrm{Multivariate-Normal}\left(\mathbf{0}, \Sigma^{-1} \right) \\
Y_{ij} &\sim \textrm{Poisson}( \exp\{X_{i:}\beta_j + w_{ij}\} )
\end{align*}
Here $\mathbf{0}$ and $\Sigma^{-1}$ are the $1 \times o$ zero mean vector and $o \times o$ precision matrix of the multivariate normal, and  $w$ is an $n \times o$ latent abundance matrix. The coefficient matrix, $\beta$, is $p \times o$ such that $\beta_{ij}$ denotes predictor $i$'s coefficient for taxon $j$. 

The log-posterior of this model is given by 
\begin{align}
 \sum_{j=1}^{o} \sum_{i = 1}^{n} & \left[  y_{ij}( x_{i:}\beta_{:j} + w_{ij} ) -  \exp\{x_{i:}\beta_{:j} + w_{ij} \} \right] + \frac{n}{2} \log|\Sigma^{-1}| - \frac{n}{2}\textrm{tr}\left(S(w)\Sigma^{-1} \right)
\end{align}
where $S(w) = w^{T}w/n$ is the empirical covariance matrix of $w$. 

Intuitively, the columns of $w$ are adjusted, ``residual'' abundance measurements of each bacteria, after controlling for confounding predictors in $X$. Assuming all relevant confounding covariates are indeed included in $X$, the only signal that remains in these residuals must arise from interactions between the bacteria being modeled. Therefore, we wish to model direct interactions, or equivalently, conditional independences at the level of these latent abundances, rather than the observed counts. Recall if $\Sigma^{-1}_{ij} = 0$, then $w_{:i}$ and $w_{:j}$ are conditionally independent, and so too are $Y_{i:}$ and $Y_{j:}$ since the probability density of $Y_{:k}$ given $w_{:k}$ is completely determined. Thus, assuming a correct model, $\Sigma^{-1}_{ij} = 0$ is sufficient to conclude that bacteria $i$ and bacteria $j$ do not interact, and are conditionally independent given all other bacteria. Similarly, if $\Sigma^{-1}_{ij} \ne 0$, we would conclude that bacteria $i$ and bacteria $j$ do directly interact.  

To appreciate the degree of coupling between two bacteria we must normalize $\Sigma^{-1}_{ij}$ to $\Sigma^{-1}_{ii}$ and $\Sigma^{-1}_{jj}$. A large $|\Sigma^{-1}_{ij}|$ need not be indicative of a strong coupling if, for example,  $\Sigma^{-1}_{jj}$ and $\Sigma^{-1}_{ii}$ -- the conditional variance of bacteria $i$ and $j$ given all others -- are much larger. Therefore, in subsequent results and visualizations we consider a transformation of $\Sigma^{-1}$ to its partial correlation matrix, $P$, whose entries are specified as $P_{ij} = -\Sigma_{ij}/\sqrt{\Sigma^{-1}_{ii}\Sigma^{-1}_{jj}}$. 

Finally, we wish to have an estimate of an interaction network that not only well explains the correlated count data we observe, but also does so parsimoniously, in a manner that maximizes the number of correct hypotheses and minimizes the nubmer of false ones that lead to wasted testing effort. Toward this end, we impose an adjustable $\ell_1$-penalty on the entries of the precision matrix during optimization, which encourages the precision matrix to be sparse. Importantly, from a Bayesian perspective, the $\ell_1$ penalty can be seen as a zero-mean Laplace distribution (with parameter $\lambda$) over the model parameter it is regularizing.

\subsubsection*{Model learning}
The $\ell_1$-penalized log-posterior, modulo unnecessary constants, is given by 
\begin{align*}
\argmax_{\beta, w, \mu, \Sigma^{-1}} \sum_{j=1}^{o} \sum_{i = 1}^{n} & \left[  y_{ij}( x_{i:}\beta_{:j} + w_{ij} ) -  \exp\{x_{i:}\beta_{:j} + w_{ij} \} \right] \\ 
& + \frac{n}{2} \log|\Sigma^{-1}| - \frac{n}{2}\textrm{tr}\left(S(w)\Sigma^{-1} \right) - \frac{\lambda n}{2}|| \Sigma^{-1} ||_1  + o^2 \log\left( \frac{n\lambda}{4} \right) \numberthis \label{l1logpost}
\end{align*}
where $\lambda$ is a tuning parameter, and $||\cdot||_1$ denotes the $\ell_1$-norm, which for a matrix $M$ equals $\sum_i\sum_j |M_{ij}|$. Note we have presented the $\ell_1$ penalty as a Laplace distribution with parameter $2/(n\lambda)$. In other words, $f(\Sigma^{-1}_{ij} | 2/(n\lambda)) =  n\lambda \exp( -n\lambda|\Sigma^{-1}_{ij}|/2)/4$. 

We optimize this objective using an iterative conditional modes algorithm in which parameters are sequentially updated to their mode value given current estimates of the remaining parameters \cite{Besag1986}. Given an estimates of $w$, $\mu$, and $\Sigma^{-1}$, the conditional objective for $\beta$ is given by, 
\[
\argmax_{\beta} \sum_{j=1}^{o} \sum_{i = 1}^{n} \left[  y_{ij}( x_{i:}\beta_{:j} + \hat{w}_{ij} ) -  \exp\{x_{i:}\beta_{:j} + \hat{w}_{ij} \} \right] 
\]
This is efficiently and uniquely optimized by setting $\beta_{:k}$ to the solution of the Poisson regression of $Y_{:k}$ onto $X$ using a log-link and $w_{:k}$ as an offset, for all $k \in \{1, 2, \ldots, o\}$.

Given estimates for $\beta$, $\Sigma^{-1}$, and $\mu$, the conditional objective for $w$ is given by 
\begin{align*}
 \argmax_{w} \sum_{j=1}^{o} \sum_{i = 1}^{n} & \left[  y_{ij} w_{ij}  -  \exp\{x_{i:}\hat{\beta}_{:j} + w_{ij} \} \right] - \frac{n}{2}\textrm{tr}\left(S(w)\hat{\Sigma}^{-1} \right) 
\end{align*}
Each row of $w$ is independent of all other rows in this objective and can therefore be updated separately. To obtain the conditional update for $w_{i:}$, we apply Newton-Raphson. The gradient vector, $g_i$, and Hessian, $H_i$, are given by 
\[
g_i = y_{i:} - \exp\{x_{i:}\hat{\beta}_{:j} + w_{ij} \} - (w_{i:} - \hat{\mu})\hat{\Sigma}^{-1} \quad H_i = -\hat{\Sigma}^{-1} - \textrm{diag}(\exp\{ x_{i:}\hat{\beta} + w_{i:}\} )
\]
Because $\hat{\Sigma}^{-1}$ is positive-definite and $\exp\{ x_{i:}\hat{\beta} + w_{i:}\} > 0$ for all components, $H_i$ is always negative-definite. Thus, the conditional update for $w_{i:}$ is a unique solution.

Given $\beta$, $w$, and $\mu$, the conditional objective for $\Sigma^{-1}$ is given by,
\[
\argmax_{\Sigma^{-1}} \log|\Sigma^{-1}| - \textrm{tr}\left(S(w)\Sigma^{-1} \right) - \lambda || \Sigma^{-1} ||_1
\]
which is convex, and efficiently optimized using the graphical lasso \cite{Friedman2008}. 

\subsubsection*{Model initialization}

In a manner similar to our conditional update for $\beta$, we initialize $\beta_{:k}$ to be the solution of the Poisson regression of $Y_{:k}$ onto $X$ using a log-link, but with no offset,  for all $k \in \{1, 2, \ldots, o\}$. Given this $\beta$, the predicted mean of the associated Poisson distribution is given by $\mathbb{E}(Y_{ij}|X) = \exp(X_{i:}\beta_j)$. Note, however, in the original formulation of our model, we have $\mathbb{E}(Y_{ij}|X) = \exp(X_{i:}\beta_j + w_{ij})$. This suggests a natural initialization for $w_{ij}$: $w_{ij} = \log(Y_{ij}) - X_{i:}\beta_j$. To complete the initialization, we set $\Sigma^{-1}$ to be the generalized pseudoinverse of $S(w)$ -- a numerically stable estimate of the precision matrix of $w$. The rationale behind this initialization is consistent with the previously presented intuitions underlying the components of each model, and in practice leads to quick convergence.

\subsubsection*{Model selection}


The $\ell_1$ tuning parameter, $\lambda$, is a hyperparameter that must be set before the model can be learned. In supervised learning, cross validation is a popular method used to set such penalties. In our model, however, $w$ is a sample specific parameter that consequently must be estimated for held out data before prediction error can be evaluated. This breaks the independence assumption between training data and test data, and in general, results in poor or undeterminable model selection; less penalizing (smaller) values of $\lambda$ tend to always produce statistically lower test-set prediction error, because $w$ is allowed to ``adapt'' to the test set samples.

Instead of cross validation, we assume only for the purpose of selecting a value for $\lambda$ that there is a joint distribution between between $\lambda$ and the remaining parameters, in which $\lambda$ has an improper flat prior (the prior probability density of $\lambda$ always equals 1). Then, differentiating Equation \ref{l1logpost}, setting equal to 0, and solving for $\lambda$, gives us $\hat{\lambda} = 2o^2/\left( n ||\Sigma^{-1}_{\textrm{init}}|| \right)$, which is the value of $\lambda$ we use throughout the optimization. Here, $\Sigma^{-1}_{\textrm{init}}$ is our initial estimate of $\Sigma^{-1}$ and is obtained as described in the previous section.

We note here a qualitative connection to empirical Bayes inference, in which hyperparameter values are set to be the maximizers of the marginal likelihood  -- the probability density of the data given only the hyperparameters. In effect, emperical Bayes calculates the expected posterior density by averaging over model parameters, and then chooses the hyperparameter value that maximizes it. In our case, instead of marginalizing over parameters, we make an intelligent guess at their value, and condition on these values to set our hyperparameter $\lambda$. In both methods, hyperparameters are set in an unbiased, and objective way by looking first at the data.

\subsection*{Synthetic Experiment}

To test our model's accuracy, efficiency, and performance relative to other leading methods, we constructed a 20-node synthetic experiment composed of 100 samples. As our precision matrix, $\Sigma^{-1}$ we generated a random, 20 $\times$ 20,  85\% sparse (total of 27 non-zero, off-diagonal entries) positive-definite matrix using the \texttt{sprandsym} function in MATLAB. From $\Sigma^{-1}$ we generated latent abundances, $w$, for 100 samples using a standard multivariate normal random variable generator based on the Cholesky decomposition. We then generated two ```confounding'' covariates, $X_1$ and $X_2$. $X_1$ was a vector of 100 independent and identically distributed $\textrm{Normal(4,1)}$ random variables. $X_2$ was 100-long vector where the first 50 entries equaled 1 and the last 50 equaled 0. The weights, $\beta_{1j}$ and $\beta_{2j}$ on each confounding covariate, were set to be -0.5 and 6, respectively,  for all nodes (i.e. for all $j \in \{1, 2, \ldots, 20\}$). These coefficient values were chosen such that the combined effect size of these confounding covariates on the response was 3 times larger than the effect size of the latent abundances, or equivalently, the contribution of the interactions encoded in the precision matrix. The 100 $\times$ 20 response matrix, $Y$, was generated according to $Y_{ij} \sim \textrm{Poisson}(X_{i1}\beta_{1j} + X_{i2}\beta_{2j} + w_{ij})$. Finally, for the same precision matrix, we generated 20 replicate response matrices in this manner. 

We applied our model to the 100 $\times$ 20 synthetically generated response matrix $Y$, and entered the counfounding covariates, $X_1$ and $X_2$ as predictors. We also applied SparCC and the graphical lasso (glasso) to illustrate the performance of a state-of-the-art correlation based method and a widely used method for inferring graphical models, respectively. While we applied SparCC to $Y$ only, we ran glasso on a matrix composed of the column-wise concatenation of $Y$ and $X$, effectively learning a joint precision matrix over nodes represented in $Y$ and the covariates in $X$. Applying glasso in this manner allowed it to account for the confounding predictors in $X$. To compare the glasso learned precision matrix to the true precision matrix, we use only the 20 $\times$ 20 subset matrix corresponding to the nodes represented in $Y$. The $\ell_1$ tuning parameter for glasso was chosen by cross-validation, where the selection criterion was the test-set log-likelihood.

\subsection*{Artificial Community Experiment}

To test the model with real data, we constructed a 9 member artificial community composed of \emph{Escherichia coli} (a putative negative root colonization control) and 8 other bacterial strains originally isolated from \emph{Arabidopsis thaliana} roots grown in two wild soils \cite{Lundberg2012}. These 8 isolates were chosen based on their potential to confer beneficial phenotypes to the host (unpublished data) and to maximize phylogenetic diversity. Into each of 94 2.5-inch-square pots pots filled with 100mL of a 2$\times$ autoclaved, calcined clay soil substrate, we inoculated the 9 isolates in varying relative abundances in order to perturb their underlying interaction structure. For all pots, all strains were present, but ranged in input abundance from 0.5-50\%. 

To each of these inoculated pots, we carefully and asceptically transferred a single sterilely grown Col-0 \emph{A. thaliana} seedling. Pots were spatially randomized and placed in growth chambers providing short days of 8 h light at 21$^{\circ}$C and 16 h dark at 18$^{\circ}$C. The plants were allowed to grow for four weeks, after which we harvested their roots and for each, performed 16S profiling (includes DNA extraction, PCR, and sequencing) of the V4 variable region. To quantify the relative amount of each input bacterium, sequencing reads were demultiplexed, quality-filtered, adjusted to ConSeqs if applicable (see Batch B processing below), and then mapped using the Burrows Wheeler Aligner \cite{Li2010}, to a previously constructed sequence database of each isolate's V4 sequence. Mapped ConSeqs or reads to a given isolate in a given sample were counted and subsequently assembled into a 94-samples $\times$ 9-isolates count matrix.

While all 94 samples were harvested over two days, they were thereafter processed in two batches, A (52 samples) and B (42 samples), approximately 4 months apart. Batch B samples were 16S profiled using the method described in \cite{Lundberg2013}. This PCR method partially adjusts for sequencing error and PCR bias by tagging all input DNA template molecules with a unique 13-mer molecular tag prior to PCR. After sequencing, this tag is then used to informatically collapse all tag-sharing amplicon reads into a single consensus sequence, or ConSeq. Batch A samples were 16S profiled by using a more traditional PCR. Having two distinct sample sets, each processed using different protocols, allowed us to assess our model's ability to statistically account for batch effects when inferring the interaction network of our 9 member community.

\subsection*{\emph{In vitro} Coplating Validation Experiments}

To test predicted interactions from our artificial community experiment, we grew liquid cultures of predicted interactors and non-interactors to OD 1 in 2XYT liquid media. We then coplated 6 5 uL dots of predicted interactors and non-interactors on King's B media agar plates, either 1 cm apart (3 dots each) or 12 cm (3 dots each) apart on the same plate. We then examined each strain for growth enhancement or restriction that was specific to its proximity to the potential interactor it was tested against.

\section*{Results}

\subsection*{Synthetic experiment}
\begin{figure*}[!tp]
\centerline{\includegraphics[width=\textwidth,height=\textheight,keepaspectratio,trim=0cm 55cm 0cm 0cm,clip]{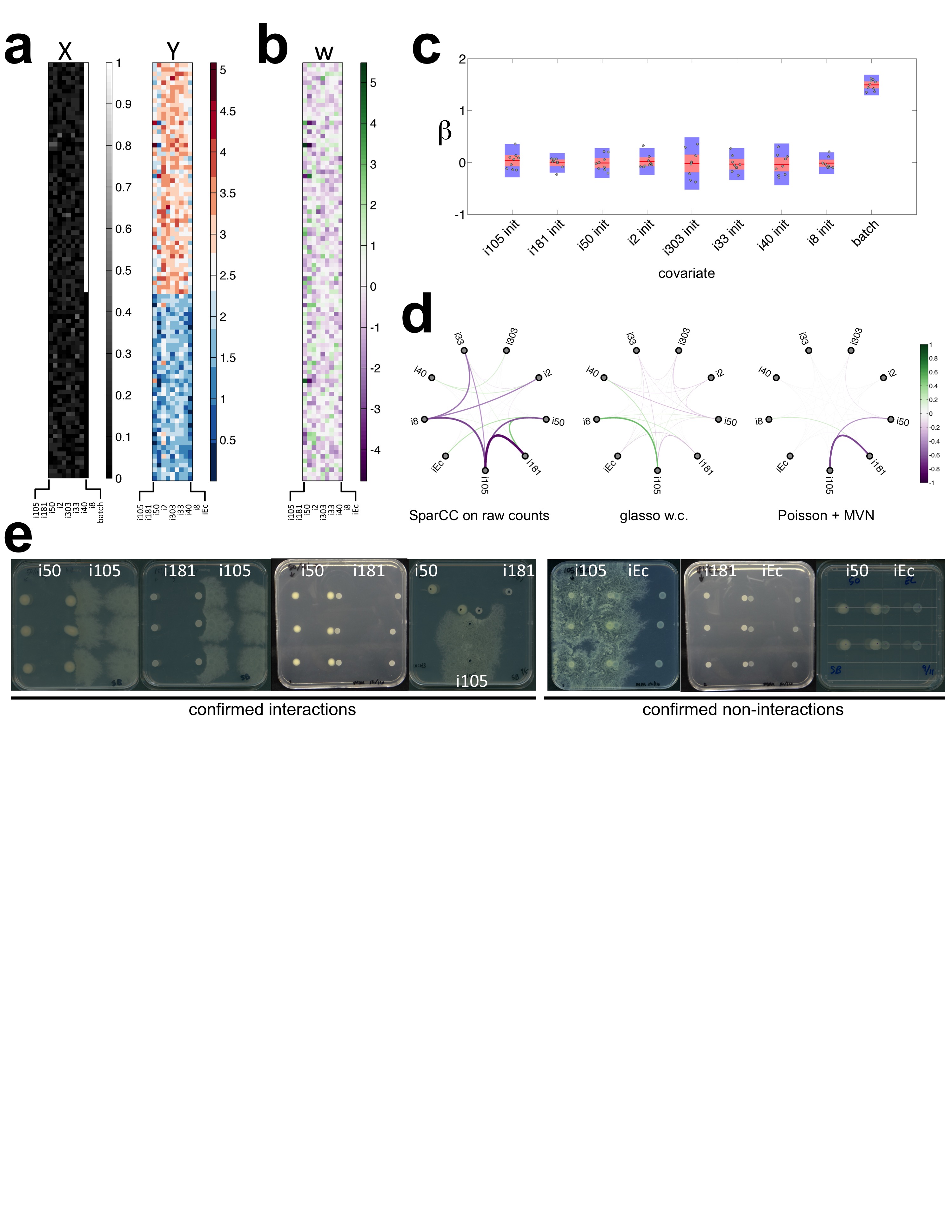}}
\caption{\textbf{The Poisson-multivariate normal hierarchical model outperforms SparCC and glasso in a synthetic experiment.} a) Frobenius norm of the difference between the partial correlation transformed true precision matrix and the estimated precision matrix for each method. The graphical lasso was run jointly over all response variables and covariates, and is therefore suffixed with ``w.c.'' (with covariates). Shaded blue bands represent 2$\times$ standard deviation and shaded red bands represent 2$\times$ standard error. b) False discovery rate of each method as a function of the number of magnitude-ordered edges called significant. The solid thick line illustrates the average FDR curve across all replicates. The shaded bands illustrate the 5$^{th}$ and 95$^{th}$ percentile FDR curve considering all replicates. Network representations of the c) true partial correlation transformed precision matrix d) correlation matrix outputted by SparCC, e) partial correlation transformed precision matrix outputted by glasso w.c. and f) partial correlation transformed precision matrix outputted by our Poisson-multivariate normal hierarchical model.}
\label{synthetic}
\end{figure*}

Figure \ref{synthetic} illustrates performances for the three methods.  With the exception of SparCC, Figure \ref{synthetic}a illustrates the Frobenious norm of the difference between the partial correlation transformations of the true precision matrix and the estimated one. The Frobenious norm, also called the Euclidean norm, is equivalent to an entry-wise Euclidean distance between two matrices, and is therefore a measure of the closeness two matrices when computed on their entry-wise difference. For SparCC, the difference is caluclated between the true partial correlation matrix and the estimated correlation matrix. 

SparCC's correlation matrix is the most different from the true partial correlation matrix, followed by glasso with covariates (w.c.) entered as variables. Our Poisson-multivariate normal hierarchical model performs the best, and interestingly, is the most consistent across replicates than the other methods.

Figure \ref{synthetic}b illustrates a complimentary measure of accuracy, the false discovery rate, which is defined to be the number of falsely non-zero edges inferred divided by the total number of non-zero edges inferred. More specifically, Figure \ref{synthetic}b illustrates FDR as a function of the number of edges (ordered by descending magnitude) called significant. 

Here again we see SparCC has the least desirable performance with and FDR curve that nearly majorizes glasso w.c. and completely majorizes our method. The graphical lasso has the next most desirable FDR curve, but still has 3 to 4 false discoveries in the top 10 non-zero edges. Our method outperforms the other two and incurs nearly 0 false discoveries in the top 10 (in magnitude) edges it discovers across all replicates. 

Figure \ref{synthetic}d, e, and f illustrate network representations for the average (across all replicates) correlation or partial correlation matrix learned by each method. Figure \ref{synthetic}c provides the network representation of the true partial correlation matrix used in this synthetic experiment.

These networks visually support previous claims. The network produced by SparCC is not sparse and is visually most distant from the true network. The glasso w.c. method is considerably more sparse, and seems to recover several of the of the top positive edges. Our method's network is visually closest to the true network and recovers considerably more of the top edges. However, it also detects them with less magnitude.

\subsection*{Artificial Community Experiment}
\begin{figure*}[!tp]
\centerline{\includegraphics[width=\textwidth,height=\textheight,keepaspectratio,trim=0cm 60cm 0cm 0cm,clip]{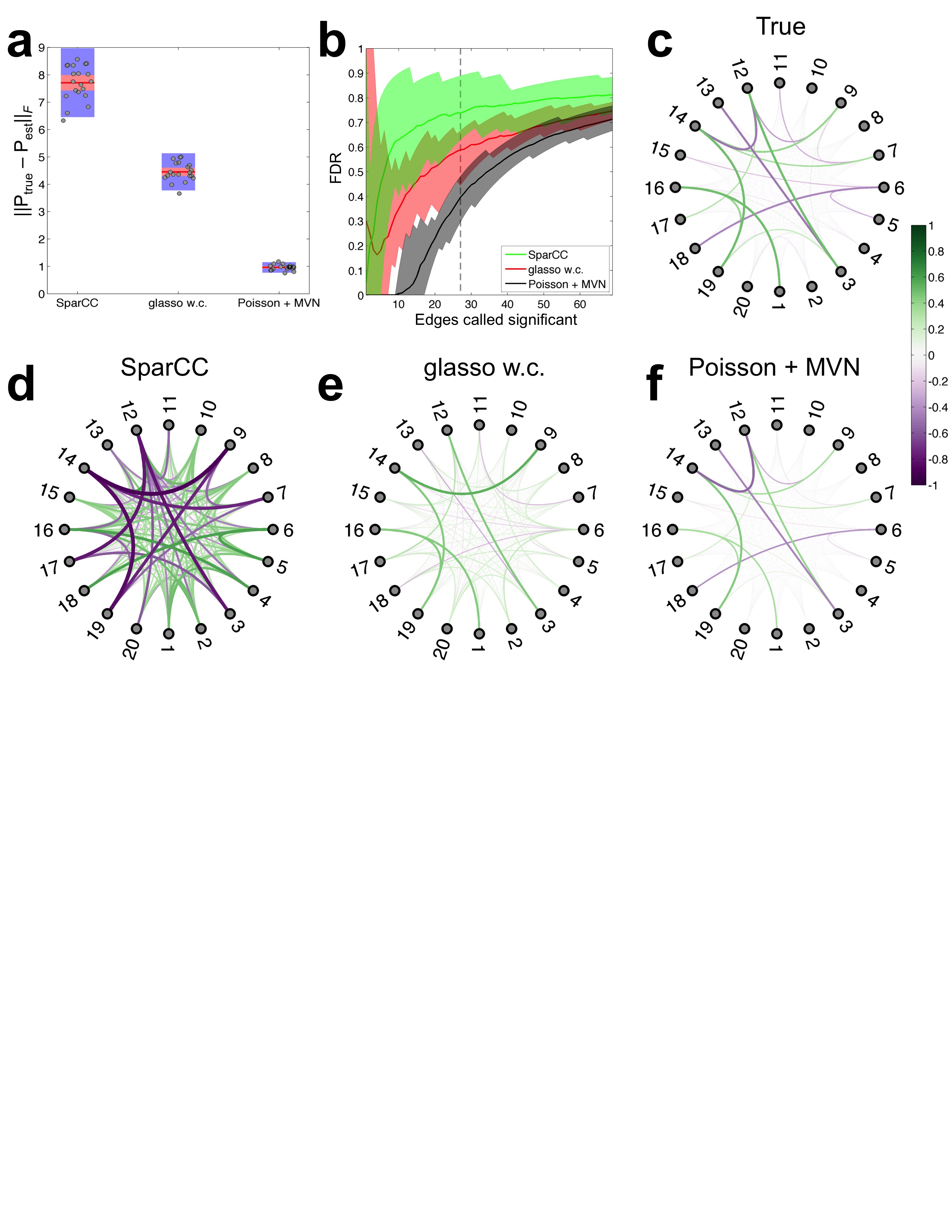}}
\caption{\textbf{Re-colonization and isolate-isolate interaction results from the 9 member synthetic community.}  a) Design (left) and response (right) matrices. The design matrix was composed of a binary vector idicating processing batch and the relative input abundances of each input isolate except \emph{E. coli} (to preserve rank). Prior to running the model, the design matrix was standardized so that coefficients on each variable could be directly comparable. Response matrix illustrates raw-counts on a $\log_{10}$ scale.  b) Latent abundance matrix, $w$, inferred from our model. These latent abundances are read counts of each bacteria ``adjusted'' for the covariates encoded in $X$.  c) Dotted box-plot illustrating the effect size of each predictor on each isolate, presented as a single dot for each predictor. Purple bands illustrate 2$\times$ standard deviation and red bands illustrate 2$\times$ standard error. d) Network visualizations of correlation matrices outputted from SparCC run on raw counts (left), partial correlation transformation of the precision matrix outputted by glasso w.c. (middle), and the partial correlation matrix obtained from the sparse precision matrix inferred from our mode (right)l. e) Interaction and non-interaction predictions tested \emph{in vitro} on agar plates. The rightmost co-plating among the ``confirmed interactions'' attempts to directly test the conditional independence structure of the (i181, i50, and i105) triad.}
\label{syncom}
\end{figure*}

We applied our model to the 94 root-samples $\times$ 9 isolates count matrix. Starting input abundances and processing batch (Figure \ref{syncom}a, left) were entered as covariates in our model.  Prior to running the model, the design matrix was standardized so that coefficients on each variable could be directly comparable.

In examining the response matrix (Figure \ref{syncom}, right) we notice a clear difference in the number of counts between Batch A samples and Batch B samples. This is due to the molecule tag correction that was available and applied to Batch B samples. The molecule tag correction collapses all reads sharing the same molecule tag into a single ConSeq -- a representative of the original template molecule of DNA, prior to PCR. However, in examining the latent abundances, $w$, (Figure \ref{syncom}b) we notice the model has successfully adjusted for these effects. As we would also expect, Figure \ref{syncom}c illustrates that the learned effect size of the batch variable is an influential predictor of bacterial read counts, more so than the starting abundance of each bacteria. 

Interestingly, in further scrutinizing Figure \ref{syncom}b we notice an interesting correlation in the latent abundances of i181 and i105, and to a lesser extent between i50 and i105. As a corallary, the latent abundances of i181 and i50 are also correlated. These correlations are suggestive of direct interaction relationships among these three bacteria, but a number of direct interaction structures could explain these correlations.

Figure \ref{syncom}d illustrates network visualizations of either correlation matrices outputted from SparCC (left), or partial correlation transformed precision matricess outputted by glasso with covariates (w.c.) entered as variables, or by our model. SparCC applied to the raw response matrix suggests a number of negative correlations that include all community members except i303. Interestingly, SparCC, which operates on log-ratios of the bacterial counts, seems immune the positive correlations among the community members one would expect to arise due to the processing batch effect. The graphical lasso with covariates entered as variables affords the simplest model, and only suggests a positive interaction between i8 and i105. 


The precision matrix our model infers is sparse, containing only two edges -- one between i105 and i181, and another between i105 and i50. The network representation of the partial correlation matrix of our precision matrix reveals a strong predicted direct antagonism between i181 and i105 and also to a lesser extent between i105 and i50. Note that the model does not predict any interaction between i181 and i50.

\emph{In vitro} co-plating experiments corroborate the model's predictions exactly in direction and also semi-quantitatively (Figure \ref{syncom}e). In particular, they show that (i105, i181) and (i105, i50) are, indeed, antagonistic interaction pairs, and moreover, that i181 and i50 are the inhibitors. Additionally, the (i181, i105) inhibition appears more pronounced than the (i181, i50) inhibition, just as the model suggests. The model also predicts conditional independence of i50 and i181 given i105. Indeed, the inward facing edges of the i181 and i50 colonies do not appear deformed in either the paired co-plating the triangular co-plating, and therefore suggest a non-interaction. Note that naively interpreting the SparCC network edges as evidence of direct interaction would falsely lead one to conclude that a direct, positive interaction exists between i50 and i181. 

Finally, note that our model predicts that i181, i50, and i105 do not interact with many of the other community members. As support for this prediction, we see that neither i105, i50, nor i181 interact with iEc. 

\section*{Discussion}

We demonstrated our Poisson-multivariate normal hierarchical model can infer true, direct microbe-microbe interactions in synthetic and real data. Proper modeling of confounding predictors is necessary to detect the (i105, i181) and (i105, i50) interactions. Though not illustrated for brevity, without controlling for processing batch, the model detects a large number of positive interactions, none of which are supported in our co-plating experiments. 

While SparCC is capable of detecting the top correlations between directly interacting members, it is unable to successfully resolve the correct conditional independence structure among them, despite its intention to produce a sparse network. Though the graphical lasso can infer direct interactions, its inability to properly model covariates or count based abundance measurements greatly reduces its utility in metagenomic sequencing experiments. Finally, though not derived for brevity, we note that the Poisson-multivariate normal model has as flexible as a mean-variance relationship as a negative binomial model, and can therefore readly handle overdispersion. Intuitively, it's modeling the Poisson mean as a log-normal random variable affords this flexibility.

We conclude that our method provides a structured, accurate, and distributionally reasonable way of modeling correlated count based random variables and capturing direct interactions among them.

\bibliography{mclb_ref}

\end{document}